\documentclass[twocolumn, longauthor, tighten]{aastex701}

\usepackage{xspace}

\newcommand{\target}{\object{COSMOSVLA J100019.21+021337.2}\xspace}
\shorttitle{JWST-Dark Radio Source}
\shortauthors{M. Li}

\begin{document}

\title{Something Bright at the Edge of Everything: A Uniquely JWST-Dark Radio Source in COSMOS}

\author[0000-0001-6251-649X]{Mingyu Li}
\affiliation{Department of Astronomy, Tsinghua University, Beijing 100084, China}
\email{lmytime@hotmail.com}


\begin{abstract}
For decades, astronomers have been searching for bright radio sources deep into the epoch of reionization (EoR). The most distant, powerful radio sources are expected to reside in heavily dust-obscured galaxies, exceedingly faint at optical and infrared wavelengths. Motivated by this, I systematically cross-match radio and JWST source catalogs in the COSMOS field and identify a uniquely JWST-dark radio source: the only object undetected in every JWST band, yet clearly detected in radio data from LOFAR 144 MHz to the VLA 3 GHz. The source is only marginally resolved and shows a steep, unbroken radio spectrum, while remaining undetected in all available HST, JWST, Chandra, Herschel, and ALMA imaging. It may represent an extremely dust-obscured radio-loud source at cosmic dawn, or alternatively a detached radio lobe whose host galaxy lies elsewhere. In either case, it highlights the new discovery space at the intersection of deep radio surveys and JWST imaging.
\end{abstract}


\section{Introduction}

Powerful radio galaxies are among the most extreme objects in the Universe, hosting the most massive black holes and residing in the most overdense environments at high redshift \citep{Miley2008A&ARv..15...67M}.
Beyond their role in galaxy evolution, high-redshift radio galaxies (HzRGs) at $z>6$ can serve as the most promising background sources for detecting the ``21~cm forest'' — \ion{H}{1} absorption imprinted on bright radio continuum by the neutral intergalactic medium during reionization \citep{Carilli2002ApJ...577...22C}.
This application is particularly compelling: even a small number of bright radio sources at cosmic dawn would transform the statistical power of 21~cm forest experiments with current and future low-frequency radio facilities.

\begin{figure*}[t]
    \centering
    \includegraphics[width=\linewidth]{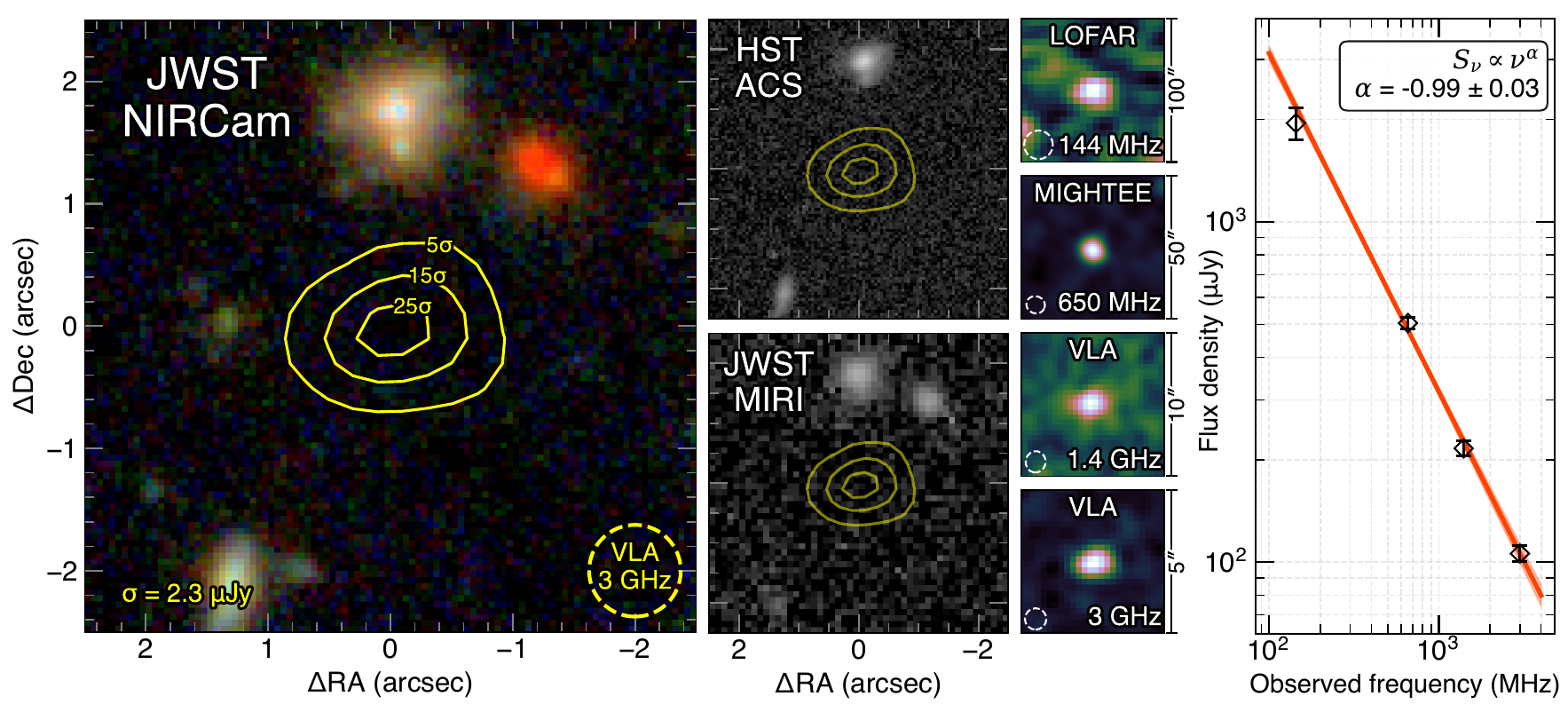}
    \includegraphics[width=\linewidth]{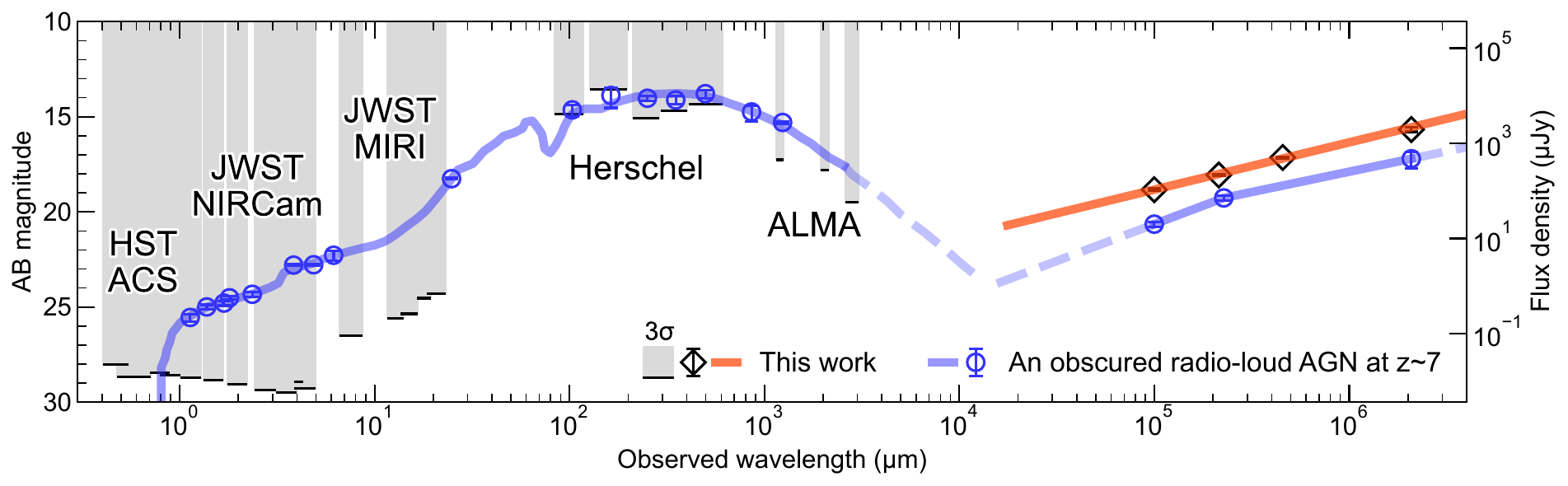}
    \caption{Multi-wavelength observations of the JWST-dark radio source. \textit{Top Left}: JWST NIRCam color image (blue: F090W+F115W+F150W; green: F200W+F277W; red: F356W+F410M+F444W) with VLA 3 GHz contours overlaid at $5,15, 25\sigma$ in yellow, where $\sigma=2.3~\mu$Jy\,beam$^{-1}$. \textit{Top Middle}: the stacked HST ACS (F475W+F606W+F814W) image, JWST MIRI image (F770W+F1280W+F1500W+F1800W+F2100W), and radio imaging data from 144~MHz to 3~GHz with dashed circles indicating the corresponding beam size. \textit{Top Right}: radio spectral fit, yielding $\alpha=-0.99\pm0.03$ for $S_\nu\propto\nu^\alpha$. \textit{Bottom}: Photometric measurements and spectral energy distribution. Gray shadows show the $3\sigma$ upper limits from HST, JWST, Herschel, and ALMA, together with the radio detections denoted by black diamonds. Blue circles and lines represent a model SED of a heavily obscured radio-loud AGN \citep{Endsley2023MNRAS.520.4609E} at $z\sim7$ for comparison.}
    \label{fig}
\end{figure*}

The search for bright radio sources deep into the EoR has been ongoing for decades, primarily by selecting ultra-steep-spectrum (USS) radio sources, reaching spectroscopic confirmations up to $z\sim5$ \citep[e.g.,][]{vanBreugel1999ApJ...518L..61V}.
In fainter radio regimes, the redshift frontier has been pushed to $z\sim7$ \citep[e.g.,][]{ Endsley2023MNRAS.520.4609E, Banados2025NatAs...9..293B}.
Despite these advances, the overall population of radio sources at $z>6$ remains poorly characterized and has yet to be fully probed deep into the EoR.
At $z\gtrsim7$, spectroscopic confirmation is exceptionally challenging due to the heavy dust obscuration.
Deep JWST/NIRCam imaging now can reveal counterparts for nearly all radio sources \citep[e.g.,][]{Gentile2025A&A...697A..46G}.
Consequently, the rare radio sources without JWST counterparts represent the most promising candidates for the highest-redshift radio galaxies.
Motivated by this realization, I perform a systematic cross-match between radio and JWST source catalogs in the COSMOS field to identify any JWST-dark radio sources.

\section{A JWST-Dark Radio Source}
I begin with the complete sample of all sources detected at $>5\sigma$ significance in the 3 GHz VLA catalog \citep{Smolcic2017A&A...602A...1S} that lie within the JWST COSMOS-Web footprint \citep{Casey2023ApJ...954...31C}, yielding a total of 2744 sources.
A coordinate cross-match within a 1-arcsec radius to the COSMOS2025 JWST source catalog \citep{Shuntov2025A&A...704A.339S} identifies 117 unmatched objects.
After visual inspection of these 117 sources, 43 could be transient or spurious sources detected only at 3 GHz with faint fluxes and without counterparts at other radio frequencies, including the VLA at 1.4 GHz \citep{Schinnerer2007ApJS..172...46S}, the superMIGHTEE at 650 MHz \citep{Lal2025ApJ...991....9L}, and the LOFAR two-metre sky survey (LOTSS) at 144 MHz \citep{Shimwell2026A&A...707A.198S}.
The remaining 73 sources are linked to a galaxy through extended radio emission or actually have JWST counterparts, but were overlooked by the COSMOS2025 catalog due to adjacent bright sources.
Among the 117 unmatched radio sources, only one (\href{https://cosmos2025.iap.fr/fitsmap/?ra=150.0800420&dec=2.2269830&zoom=10}{\target}) remains without a clear counterpart in any JWST band.

The source has integrated flux densities of $1950 \pm 210$~$\mu$Jy, $504 \pm 19$~$\mu$Jy, $216 \pm 11$~$\mu$Jy, and $106 \pm 6$~$\mu$Jy at 144~MHz, 650~MHz, 1.4~GHz, and 3~GHz, respectively.
A power-law fit yields $\alpha = -0.99 \pm 0.03$ (reduced $\chi^2 = 1.9$), a steep synchrotron spectrum with no significant evidence for spectral curvature or low-frequency turnover across the observed range.
At 3~GHz, the source is marginally resolved by the VLA ($0.75''$ synthesized beam), well fitted by a single Gaussian component with FWHM of 1.1\arcsec $\times$ 0.8\arcsec, but no clear double-lobed or core-jet morphology is resolvable at this resolution.
The source is undetected in HST/ACS, JWST/NIRCam, and JWST/MIRI images from COSMOS-Web \citep[JWST-GO-1727,][]{Casey2023ApJ...954...31C}, PRIMER (JWST-GO-1837), COSMOS-3D (JWST-GO-5893), and MINERVA \citep[JWST-GO-7814,][]{Muzzin2025arXiv250719706M}.
The HST and JWST images presented herein are retrieved from the Dawn JWST Archive (\href{https://dawn-cph.github.io/dja/}{DJA}), which performs the data reduction using the \href{https://github.com/gbrammer/grizli}{\texttt{grizli}} pipeline \citep{Brammer2023zndo...8370018B}.
The object is undetected in Chandra X-ray \citep[depth of $2.2 \times 10^{-16}$\,erg\,s$^{-1}$\,cm$^{-2}$ in the 0.5--2\,keV band;][]{Civano2016ApJ...819...62C}, Herschel far-infrared/sub-mm imaging \citep{Oliver2012MNRAS.424.1614O}, and in available ALMA mm continuum images, including CHAMPS \citep{Zavala2026ApJ...998L..36Z}, Ex-MORA \citep{Long2026ApJ...999...47L}, and A$^3$COSMOS \citep{Adscheid2024A&A...685A...1A}.
The other multiwavelength data are shown in Fig.~\ref{fig}.

\section{Discussion}

The JWST-dark radio source could be a detached lobe or hotspot, previously jetted from a host galaxy that lies elsewhere.
It is noted that at least four galaxies are detected in the JWST image within a 2\arcsec neighborhood, which could be the host galaxy candidate.
However, it is difficult to reconcile this interpretation with the compact morphology and the lack of any extended low-surface-brightness emission, counter-lobe, or plausible core/host galaxy in the current radio imaging.

I therefore explore the interpretation that the radio emission is associated with a high-redshift dust-obscured radio-loud AGN.
Compared to an obscured hyperluminous radio-loud AGN COS-87259 with clear multi-wavelength detection at $z=6.853$ \citep{Endsley2023MNRAS.520.4609E}, \target is $\gtrsim5$ times brighter in radio, yet remains undetected from UV to mm.
This likely implies a higher redshift and/or heavier dust obscuration.
At an assumed redshift of $z=7$, the rest-frame 1.4~GHz luminosity is $\log(L_\mathrm{1.4~GHz}/\mathrm{W~Hz^{-1}})=26.1$ (flat $\Lambda$CDM cosmology with $H_0=70~\rm km~s^{-1}$ and $\Omega_m=0.3$), placing the source among the luminous radio-loud quasars at $z\gtrsim6$ \citep[e.g.,][]{Gloudemans2021A&A...656A.137G}.
The partially resolved size implies a physical scale of $\lesssim$4--6 kpc at $z=7$, typical of compact jets.
The X-ray non-detection only provides a mild constraint of $\log\rm L_{2-10}/erg~s^{-1} \lesssim 43.9$ at $z=7$, assuming a photon index of $\Gamma=1.8$.
The multi-wavelength non-detections can be interpreted by a combination of high redshift and extreme dust attenuation ($A_V \gtrsim 5-10$).
Lower redshifts would require obscuration far exceeding those of typical dusty star-forming galaxies at $z\simeq6$, most of which still show faint counterparts in deep JWST images \citep[e.g.,][]{Sun2025arXiv250606418S}.

If confirmed at $z\gtrsim7$, this object will rank among the most distant radio sources known to date.
It exemplifies the powerful discovery potential at the radio--JWST intersection for probing the extremes of galaxy and AGN evolution at the edge of the observable Universe.

\begin{acknowledgments}
The author acknowledges all the teams and individuals who contributed to the acquisition and public release of the multi-wavelength datasets that enabled this discovery.
The JWST and HST data presented in this article were obtained from the Mikulski Archive for Space Telescopes (MAST) at the Space Telescope Science Institute. The specific observations analyzed can be accessed via \dataset[doi: 10.17909/8wc0-5w17]{https://doi.org/10.17909/8wc0-5w17}.
M. Li thanks Zheng Cai, Bjorn H.C. Emonts, Xiaohui Fan, and Fengwu Sun for their feedback on this manuscript.
May the Universe not fool us, and may it truly be a radio-loud AGN at cosmic dawn.
\end{acknowledgments}





%
\facilities{HST, JWST, CXO, EVLA, ALMA, LOFAR, Herschel, MeerKAT, GMRT}





\bibliography{main}{}
\bibliographystyle{aasjournalv7}



\end{document}